\documentclass[aps,prc,reprint,superscriptaddress,twocolumn,showpacs]{revtex4-1}

\usepackage{graphicx}
\usepackage{natbib}

\begin{document}


\title{Test of statistical model cross section calculations for $\alpha$-induced reactions on $^{107}$Ag at energies of astrophysical interest}

\author{C. Yal\c{c}\i{n}}

\email[Corresponding author:]{ caner.yalcin@kocaeli.edu.tr}
\affiliation{Kocaeli University, Department of Physics, Umuttepe 41380, Kocaeli, Turkey\\}
\affiliation{Institute for Nuclear Research (MTA Atomki), H-4001 Debrecen, POB.51., Hungary\\}%

\author{Gy. Gy\"{u}rky}
\affiliation{Institute for Nuclear Research (MTA Atomki), H-4001 Debrecen, POB.51., Hungary\\}%

\author{T. Rauscher}
\affiliation{Centre for Astrophysics Research, University of Hertfordshire, Hatfield AL10 9AB, United Kingdom}
\affiliation{Department of Physics, University of Basel, 4056 Basel, Switzerland\\}

\author{G. G. Kiss} 
\altaffiliation[Present address: ]{RIKEN Nishina Center, 2-1 Hirosawa, Wako, Saitama 351-0198, Japan}
\affiliation{Institute for Nuclear Research (MTA Atomki), H-4001 Debrecen, POB.51., Hungary\\}%

\author{N. \"{O}zkan}
\author{R. T. G\"{u}ray}%
\affiliation{Kocaeli University, Department of Physics, Umuttepe 41380, Kocaeli, Turkey\\}%

\author{Z. Hal\'asz}
\affiliation{Institute for Nuclear Research (MTA Atomki), H-4001 Debrecen, POB.51., Hungary\\}%

\author{T. Sz\"{u}cs}
\altaffiliation[Present address: ]{Helmholtz-Zentrum Dresden-Rossendorf (HZDR), D-01328 Dresden, Germany}
\affiliation{Institute for Nuclear Research (MTA Atomki), H-4001 Debrecen, POB.51., Hungary\\}%

\author{Zs. F\"{u}l\"{o}p }
\author{J. Farkas}
\affiliation{Institute for Nuclear Research (MTA Atomki), H-4001 Debrecen, POB.51., Hungary\\}%

\author{Z. Korkulu}%
\affiliation{Kocaeli University, Department of Physics, Umuttepe 41380, Kocaeli, Turkey\\}%

\author{E. Somorjai}
\affiliation{Institute for Nuclear Research (MTA Atomki), H-4001 Debrecen, POB.51., Hungary\\}%

\date{\today}

\begin{abstract}
\begin{description}

\item[Background]

Astrophysical reaction rates, which are mostly derived from theoretical cross sections, are necessary input to nuclear reaction network simulations for studying the origin of $p$ nuclei. Past experiments have found a considerable difference between theoretical and experimental cross sections in some cases, especially for ($\alpha$,$\gamma$) reactions at low energy. Therefore, it is important to experimentally test theoretical cross section predictions at low, astrophysically relevant energies. 

\item[Purpose]
The aim is to measure reaction cross sections of $^{107}$Ag($\alpha$,$\gamma$)$^{111}$In and $^{107}$Ag($\alpha$,n)$^{110}$In at low energies in order to extend the experimental database for astrophysical reactions involving $\alpha$ particles towards lower mass numbers. Reaction rate predictions are very sensitive to the optical model parameters and this introduces a large uncertainty into theoretical rates involving $\alpha$ particles at low energy. We have also used Hauser-Feshbach statistical model calculations to study the origin of possible discrepancies between prediction and data.

\item[Method]
An activation technique has been used to measure the reaction cross sections at effective center of mass energies between 7.79 MeV and 12.50 MeV. Isomeric and ground state cross sections of the ($\alpha$,n) reaction were determined separately.

\item[Results]
The measured cross sections were found to be lower than theoretical predictions for the ($\alpha$,$\gamma$) reaction. Varying the calculated averaged widths in the Hauser-Feshbach model, it became evident that the data for the ($\alpha$,$\gamma$) and ($\alpha$,n) reactions can only be simultaneously reproduced when rescaling the ratio of $\gamma$- to neutron width and using an energy-dependent imaginary part in the optical $\alpha$+$^{107}$Ag potential.

\item[Conclusions]

The new data extend the range of measured charged-particle cross sections for astrophysical applications to lower mass numbers and lower energies. The modifications in the model predictions required to reproduce the present data are fully consistent with what was found in previous investigations. Thus, our results confirm the previously suggested energy-dependent modification of the optical $\alpha$+nucleus potential.

\end{description}
\end{abstract}



\maketitle

\section{Introduction}
\label{sec:intro}

The synthesis of elements heavier than iron proceeds via different processes.
The so-called $s$ and $r$ processes involve neutron capture reactions. The $s$ process is the slow neutron capture process responsible for the production of stable isotopes along the valley of beta stability in the chart of isotopes. The $r$ process is the rapid neutron capture process and approximately half of the heavy elements with mass number $A>70$ and all the actinides in the solar system are believed to have been produced by the $r$ process. These two processes were found to be unable, however, to create 35 neutron-deficient, natural isotopes between $^{74}$Se and $^{196}$Hg, which were called ``$p$ nuclei'' or ``excluded isotopes'' \cite{Burbideg57,Cameron57}. Recently, it has been shown, on the other hand, that $^{164}$Er,$^{152}$Gd and $^{180}$Ta may have large $s$-process contributions, nevertheless, and that the $\nu$-process may contribute to $^{138}$La and $^{180}$Ta (see, e.g., \cite{Rauscher2013} and references therein). The remaining $p$ nuclei are thought to be produced in the $\gamma$-process which includes combination of ($\gamma$,n), ($\gamma$,p) and ($\gamma$,$\alpha$) reactions \cite{Burbideg57,Cameron57,Woosley78,Rayet90,Cowan91,Wallerstein97,Arnould07,Baruah09,Rauscher2013}.

The $\gamma$-process nucleosynthesis is modeled by using an extended nuclear reaction network, for which -- among others -- reaction rate information of thousands of neutron, proton and $\alpha$-induced reactions as well as their inverse reactions are needed \cite{Raus00,Raus01,Raus06}. Experimental studies of reactions important in this context have been performed in recent years \cite{Fulop96,Som98,Ozkan02,Rapp02,Basunia05,Haris05,Gyu06,Ozkan07,Rapp08,Danil08,Yalcin09,Guray09,Gyurky10,Kiss11,Sauerwein11,Dillmann11,Palumbo12,Halasz12,Kiss12,Raus12,Sauerwein12,Netterdon2013,Kiss14} but despite this effort, experimental cross sections are still very scarce at astrophysically interesting, low energies. The full list of the experiments can be found in the KADoNiS $p$-process database \cite{Szucs14}. The experiments performed so far have shown that there can be considerable differences between theoretical and experimental cross sections in some cases at energies close around the Coulomb barrier. In order to get rid of this discrepancy, there is also strong effort to obtain a global $\alpha$+nucleus optical potential \cite{Morh2013_scat} via $\alpha$-elastic scattering experiments \cite{Kiss2013_scat, Gyurky2012_scat, Kiss2011_scat}. Theoretical cross sections are used in $\gamma$-process network calculations and a deficiency in reaction rates can perhaps be responsible for the failure of $\gamma$-process models in reproducing the observed $p$ isotope abundances in the mass range $150\leq A \leq 165$.  For this reason, further experimental reaction studies should be performed at astrophysical relevant energies in order to improve both the experimental cross section database and the theoretical cross section calculations.

Although $^{107}$Ag is not a $p$ nucleus and mostly produced by the $s$ and $r$ processes, in order to further test the reliability of statistical model predictions in this mass range, the $\alpha$-capture cross sections of $^{107}$Ag have been measured in the effective center of mass energy range between \mbox{7.79 MeV} and \mbox{12.50 MeV} using the activation method. These energies are close to the astrophysically relevant energy range (the Gamow window) which extends from \mbox{5.83 MeV} to \mbox{8.39 MeV} at  \mbox{3 GK} temperature  typical for the $\gamma$-process \cite{Raus10}. The results were compared with Hauser-Feshbach statistical model calculations.

Details of the experiment are given in Sec.\ \ref{sec:experiment}. The experimental results are presented in Sec.\ \ref{sec:expresults}. A comparison to statistical model calculations and a detailed discussion is given in Sec.\ \ref{sec:theory}. The final Sec.\ \ref{sec:summary} provides conclusions and a summary.

\section{Experiment}
\label{sec:experiment} 

Reaction cross sections of $^{107}$Ag($\alpha,\gamma)^{111}$In and $^{107}$Ag($\alpha,n)^{110}$In have been measured at the laboratory energies between \mbox{8.16 MeV} and \mbox{13.00 MeV}. Since the reaction products are radioactive and their half-lives are convenient, the activation method was used to determine the cross sections. Detailed information about the activation method can be found, e.g., in \cite{Ozkan02}.
	
In the case of $^{107}$Ag($\alpha$,n) the reaction product $^{110}$In has a long-lived isomeric state. The partial cross sections leading to the ground as well as the isomeric states can be determined separately owing to the different decay patterns of the two states. The decay parameters used for the analysis are summarized in Table \ref{tab:decaypar}.

\begin{table}
\caption{\label{tab:decaypar}Decay parameters of reaction products taken from \cite{Blachot09, Gurdal12}. Only the $\gamma$-transitions used for the analysis are listed.}
\begin{ruledtabular}
\begin{tabular}{cccc}
\parbox[t]{2.0cm}{\centering{Reaction}} & \parbox[t]{2.0cm}{\centering{Half-life }} & 
\parbox[t]{1.1cm}{\centering{E$_{\gamma}$}} & \parbox[t]{1.1cm}{\centering{I$_{\gamma}$}} \\
\parbox[t]{2.0cm}{\centering{}} & \parbox[t]{2.0cm}{\centering{ }} & 
\parbox[t]{1.1cm}{\centering{(keV)}} & \parbox[t]{1.1cm}{\centering{($\%$)}} \\

\hline 
$^{107}$Ag($\alpha,\gamma)^{111}$In & 2.8047 $\pm$ 0.0004 d & 171.28  & 90.7 $\pm$ 0.9 \\
                                    
$^{107}$Ag($\alpha$,n)$^{110g}$In          & 4.92 $\pm$ 0.08 h & 641.68 & 26.0 $\pm$ 0.8  \\
                                           &                   & 707.40 & 29.5 $\pm$ 1.1  \\
																					 &                   & 937.16 & 68.4 $\pm$ 1.9  \\
																					
$^{107}$Ag($\alpha$,n)$^{110m}$In          & 69.1 $\pm$ 0.5 min & 2129.40  & 2.15 $\pm$ 0.03 \\
                                           &                    & 2211.33  & 1.74 $\pm$ 0.03  \\

\end{tabular}
\end{ruledtabular}
\end{table}

\subsection{Target preparation}
Natural silver and isotopically enriched $^{107}$Ag targets were produced by vacuum evaporation onto high purity thin aluminum foils \mbox(from 1.8 $\mu$m to 2.5 $\mu$m). The backing aluminum foils were thick enough to stop the heavy reaction products. Enriched targets were produced from \mbox{99.50$\%$} isotopically enriched $^{107}$Ag available in metallic powder form (obtained from the company ISOFLEX USA, Certificate No: 47-02-107-2999). Both natural and enriched Ag metal powders were evaporated from a molybdenum crucible heated by DC current. The backing foil was placed \mbox{7 cm} above the crucible in a holder defining a circular spot with a diameter of \mbox{12 mm} on the foil for Ag deposition. 

The target thicknesses were determined with weight measurement. Before and after the evaporation the weight of the foils were measured with a precision better than \mbox{5 $\mu$g} and then from the difference the Ag areal density could be determined. Enriched and natural targets were prepared with thicknesses varying between \mbox{410 $\mu$g/cm$^{2}$} and \mbox{1042 $\mu$g/cm$^{2}$}. The targets were only irradiated once or cooled for more than 20 half-lives between two subsequent activations of the same target. Reused targets were checked by $\gamma$-measurement before the second irradiation to determine any remaining activity.   

\subsection{Activations}

The targets were irradiated with $\alpha$ beams from the cyclotron accelerator of MTA Atomki. In total thirteen irradiations were made at different energies between \mbox{E$_{lab}$ = 8.16 MeV} and \mbox{E$_{lab}$ = 13.00 MeV} laboratory energies. For \mbox{11.00 MeV} and \mbox{11.50 MeV}, two irradiations were carried out with enriched and natural targets to test systematic uncertainty related to the targets. The results were compatible with each other (see Table \ref{tab:XS}). Some energies were measured with an energy degrader foil because the cyclotron could not produce these beam energies directly (see Table \ref{tab:XS}). Aluminum and nickel foils were used as energy degraders. The thicknesses of the energy degrader foils were determined by energy loss measurement of $\alpha$ particle emitted from a $^{241}$Am source. In order to calculate thickness of the degrader foils, ThiMeT code \cite{Thimet-web} was used which takes into account energy dependence of stopping power through the degrader foil.

\begin{figure}
\resizebox{0.45\textwidth}{!}{%
\includegraphics{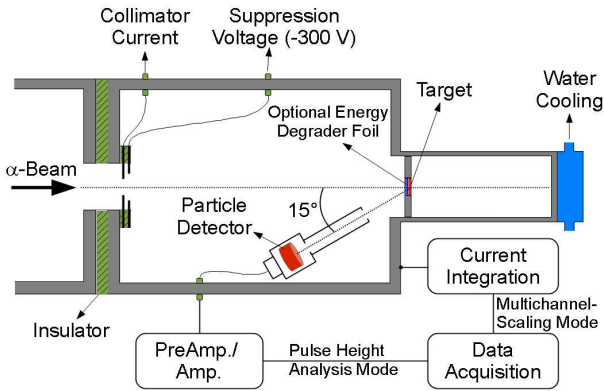}}
\caption{\label{fig:1}
(Color online) A schematic drawing of the target chamber used for the irradiations.}
\end{figure}

A diagram of the target chamber is shown in \mbox{Fig. \ref{fig:1}}.  After the last beam defining aperture the whole chamber was isolated and used as a Faraday cup to determine the number of projectiles by charge collection. A suppression voltage of \mbox{$-300$ V} was applied at the entrance of the chamber to suppress the secondary electrons. The beam current was recorded using a current integrator in multi channel scaling mode in order to take into account the possible changes in the beam current. The integrated current was recorded every 10 or \mbox{60 seconds.} 

In addition, in order to monitor target stability during the irradiation, an ion-implanted Si detector was placed into the target chamber at \mbox{165$^{\circ}$} relative to the beam direction. The elastic backscattering spectra were continuously taken and there were no substantial background peaks besides Ag and Al observed in the spectra. If there is no target deterioration then the ratio of the number of backscattered particle to those of incoming particles should be constant in time. Target stability was regularly checked and no target deterioration was observed during the irradiations. Because the target stability could not be monitored when an energy degrader foil was used, the beam current was limited to \mbox{800 nA} in these cases. This value was tested before the experiment using a natural target and found that there was no target deterioration. The beam stop was placed 10 cm behind the target from where no backscattered particles could reach the particle detector. The beam stop was directly water cooled during the irradiation. The typical current was between \mbox{150 nA} and \mbox{800 nA}. The length of irradiation was chosen in the range of \mbox{1.5 h}$-$\mbox{19.8 h} based on the longest half-life of the activation products and the expected cross section. Due to the steeply decreasing cross sections at low beam energies, longer irradiation time was applied at the lowest energies to obtain sufficient statistics.

\subsection{Gamma counting and analysis}

After each irradiation the target was taken from the reaction chamber and placed into a low-background counting
setup to measure the $^{111}$In and $^{110}$In activities produced through the $^{107}$Ag($\alpha,\gamma)^{111}$In and
$^{107}$Ag($\alpha$,n)$^{110}$In reactions, respectively. According to the actual count rate of the reaction products the target was placed at
a distance of \mbox{10 cm} or \mbox{1 cm} from the end cap of a HPGe detector having \mbox{100$\%$} relative efficiency. To reduce the room background the HPGe detector was placed into 4$\pi$ commercial \mbox{10 cm} thick lead shield with \mbox{1 mm} cadmium and \mbox{1 mm} copper layers. 

\begin{figure}
\resizebox{0.5\textwidth}{!}{%
\includegraphics{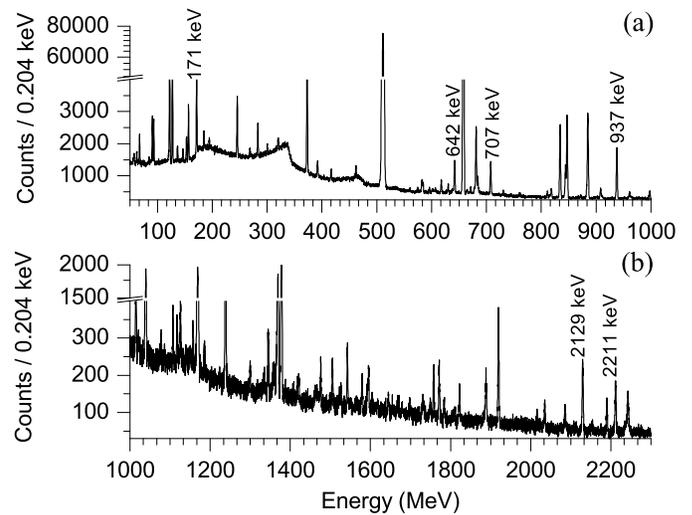}}%
\caption{\label{fig:2} 
(a) Low and (b) high energy parts of the $\gamma$ spectrum taken after an \mbox{8.7 h} irradiation of a target with a \mbox{10 MeV} $\alpha$ beam. The $\gamma$-lines used for analysis are indicated on the spectrum. The other peaks are from either laboratory background or the other $\gamma$-transitions. \mbox{(1 Channel=0.204 keV)}}
\end{figure}

As an example, \mbox{Fig. \ref{fig:2}} shows an off-line \mbox{$\gamma$-ray} spectrum taken after a \mbox{8.7 h} long irradiation
with an $\alpha$ beam of \mbox{10.00 MeV} for a counting time of \mbox{16.5 h} indicating the \mbox{$\gamma$-lines} used for cross section measurements (Table \ref{tab:decaypar}). 

Owing to the very different half-lives of the reaction products (\mbox{2.8047 d}, \mbox{4.92 h} and \mbox{69.1 min}) and the different expected cross sections, the counting periods were segmented into several parts. The $\gamma$-spectra were stored regularly in every 10 minutes near the beginning of the counting and in every 30 minutes after one hour. The ratio of the cross sections of $^{107}$Ag($\alpha$,n)$^{110}$In to $^{107}$Ag($\alpha,\gamma)^{111}$In reactions is about \mbox{30} at \mbox{10.5 MeV} and about \mbox{95} at \mbox{12.5 MeV}. At the beginning of the counting the spectra were thus dominated by the intense $\gamma$-radiations from the $^{107}$Ag($\alpha$,n)$^{110}$In decay products. The measurement of the activity of the $^{107}$Ag($\alpha,\gamma)^{111}$In reaction product was therefore started only after about six hours when the activity of at least the $^{110}$In isomeric state decreased substantially. The reaction product of the $^{107}$Ag($\alpha,\gamma)^{111}$In reaction has a short lived (\mbox{7.7 min}) isomeric state decaying completely by isomeric transition (IT) to the ground state. Starting the $\gamma$-counting for this reaction several hours after the end of the irradiation guarantees that this short lived isomer has decayed completely to the ground state and hence the total cross section can be obtained. 

The product of the $^{107}$Ag($\alpha,\gamma)^{111}$In reaction emits two strong $\gamma$-lines at \mbox{171.28 keV} and \mbox{245.35 keV} with relative intensities of \mbox{90.7$\%$} and \mbox{94.1$\%$}, respectively. But there are contributions to the \mbox{245.35 keV} peak from other decays. First, the ($\alpha,\gamma$) reaction product $^{111}$In decays to $^{111}$Cd which has an isomeric state with a half-life of \mbox{48.50 min.} This state decays with IT to the ground state and emits a \mbox{245.395 keV} $\gamma$-ray. When natural targets are used, $^{111m}$Cd is also produced by the $^{109}$Ag($\alpha$,d)$^{111}$Cd reaction above the threshold (\mbox{10.552 MeV}). A second contribution to the \mbox{245.35 keV} peak comes from the $^{112}$In isotope which is produced by the $^{109}$Ag($\alpha,n)^{112}$In reaction when natural targets are used. The energy of the gamma line is \mbox{244.8 keV} and it cannot be distinguished from the \mbox{245.35 keV} transition. There is no data for the gamma intensity of this line in literature \cite{Peker80,FRENNE89,FRENNE96} but the cross section is rather high (according to theoretical calculation with e.g. the NON-SMOKER code \cite{NON-SMOKER}, for \mbox{12.21 MeV} center of mass energy the cross section is \mbox{43.26 mb}). Because of these contributions only the \mbox{171.28 keV} $\gamma$-line was used for the cross section calculation.

\begin{figure}
\resizebox{0.5\textwidth}{!}{%
\includegraphics{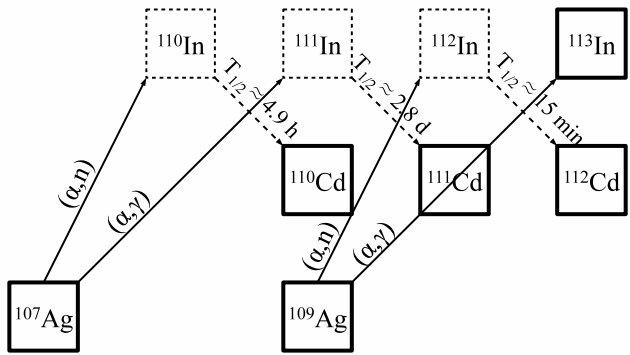}}
\caption{\label{fig:decaySch} 
Decay of the isotopes produced on natural silver targets by $\alpha$ irradiation.}
\end{figure}

In the case of the $^{107}$Ag($\alpha$,n)$^{110}$In reaction high intensity gammas at \mbox{657 keV} and \mbox{884 keV} are common for the decay of the isomeric and ground states. Therefore they were not used for the analysis. Unique gammas with high intensity for the $^{107}$In($\alpha,n)^{110g}$In reaction are at \mbox{641.68 keV}, \mbox{707.40 keV} and \mbox{937.16 keV}, and for the $^{107}$Ag($\alpha,n)^{110m}$In reaction at \mbox{2129.40 keV} and  \mbox{2211.33 keV}. These lines were chosen to determine separately the partial cross sections to the ground and isomeric states.

\subsection{Detector efficiency calibration and true coincidence summing corrections}

Absolute efficiency calibration of the detection system was done at \mbox{10 cm} detector-target distance at which the true coincidence summing effect is negligible. Calibrated $^{22}$Na, $^{54}$Mn, $^{57}$Co, $^{60}$Co, $^{65}$Zn, $^{133}$Ba, and $^{137}$Cs sources were used for the efficiency measurement. The efficiency at \mbox{171.28 keV} was determined by using a 4$^{th}$ order polynomial fitted to the calibration gamma lines in the energy range from \mbox{122.1 keV} to \mbox{1332.5 keV}. For the $^{107}$Ag($\alpha$,n)$^{110}$In case, gammas lines are located between \mbox{642.68 keV} and \mbox{2211.33 keV}. In this energy range the efficiency curve has power-law like behavior, therefore in log-log scale linear fit used between \mbox{276.4 keV} and \mbox{1332.5 keV} energies and then extrapolated to higher energies in order to find the efficiency at \mbox{2129.40 keV} and \mbox{2211.33 keV}. The validity of the linear extrapolation was checked with an uncalibrated $^{56}$Co source emitting high energy gammas. 

The efficiencies at the \mbox{1 cm} geometry used for some of the cross section measurements was determined by scaling the measured efficiencies at \mbox{10 cm}. In order to find a scaling factor for all studied $\gamma$-rays, one of the natural target was irradiated at \mbox{12.50 MeV} lab energy and counted both at \mbox{10 cm} and \mbox{1 cm}. Taking into account the lengths of the two countings and the time elapsed between them, scaling factors were determined which include both the difference in efficiency and the true coincidence summing effect in the decay of the studied In isotopes \cite{Debertin, Yalcin09}.

\section{Results and Discussion}

\subsection{Measured cross sections}
\label{sec:expresults}

\begin{table}
\caption{Measured cross sections of the $^{107}$Ag($\alpha,\gamma)^{111}$In and $^{107}$Ag($\alpha$,n)$^{110}$In reactions.}
\begin{ruledtabular}
\begin{tabular}{lccc} 

E$_{\rm beam}$ & {E$_{\rm c.m.}^{eff}$} &  \multicolumn{2}{c}{Cross section [$\mu$b]}      \\
{[MeV]}          & {[MeV]}                 &  {$^{107}$Ag($\alpha,\gamma)^{111}$In}   &{$^{107}$Ag($\alpha$,n)$^{110}$In}     \\ 
       
\hline
8.16\footnotemark[2] 									& 7.79 $\pm$ 0.08  & 0.61 $\pm$ 0.07 &                 \\
8.51\footnotemark[1]\footnotemark[2]  & 8.16 $\pm$ 0.08  & 1.52 $\pm$ 0.16 &                  \\
9.00\footnotemark[1] 									& 8.57 $\pm$ 0.06  & 2.41 $\pm$ 0.25 & 7.5  $\pm$ 1.3 \\
9.50\footnotemark[1] 									& 9.07 $\pm$ 0.06  & 5.34 $\pm$ 0.54 & 51.6 $\pm$ 4.5 \\
10.00\footnotemark[2]  								& 9.55 $\pm$ 0.09  & 12.3 $\pm$ 1.3  & 235  $\pm$ 16 \\
10.50\footnotemark[2]  								& 10.05 $\pm$ 0.09 & 26.3 $\pm$ 2.7  & 804  $\pm$ 56 \\
11.00\footnotemark[1] 								& 10.52 $\pm$ 0.07 & 50.1 $\pm$ 5.1  & 1990 $\pm$ 131 \\
11.00 																& 10.57 $\pm$ 0.07 & 52.1 $\pm$ 5.3  & 2059 $\pm$ 148 \\
11.50\footnotemark[1] 								& 11.00 $\pm$ 0.07 & 97.1 $\pm$ 9.7  & 5467 $\pm$ 354 \\
11.50 																& 11.04 $\pm$ 0.07 & 95.9 $\pm$ 10.0 & 5417 $\pm$ 365 \\
12.00 																& 11.51 $\pm$ 0.07 & 162  $\pm$ 16   & 12568 $\pm$ 803 \\
12.50 																& 12.00 $\pm$ 0.07 & 243  $\pm$ 25   & 24567 $\pm$ 1552 \\
13.00 																& 12.50 $\pm$ 0.08 & 325  $\pm$ 34   & 37066 $\pm$ 2338 \\

\end{tabular} \label{tab:XS}
\end{ruledtabular}
\footnotetext[1]{measured with enriched target.}
\footnotetext[2]{measured with an energy degrader foil.}
\end{table}

\begin{figure}
\includegraphics[angle=0,width=\columnwidth]{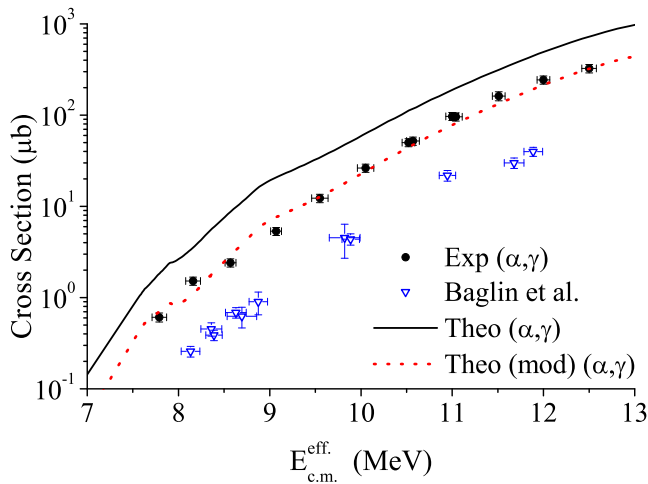}
\caption{\label{fig:ag} 
(Color online) Measured cross section of $^{107}$Ag($\alpha$,$\gamma$) compared to theory using the SMARAGD code \cite{SMARAGD} (see text for details). Previous results from Baglin \cite{Baglin2005} are also included in the figure. }
\end{figure}

\begin{figure}
\includegraphics[angle=0,width=\columnwidth]{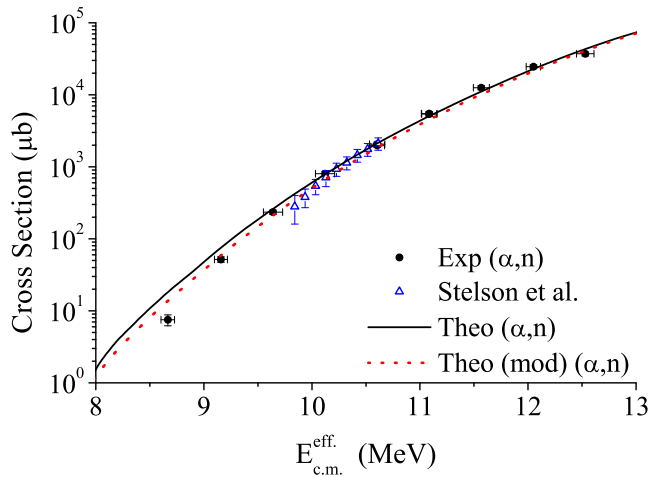}
\caption{\label{fig:an} 
(Color online) Measured cross section of $^{107}$Ag($\alpha$,n) compared to theory using the SMARAGD code \cite{SMARAGD} (see text for details). Previous results from  Stelson \cite{Stelson64} are also included in the figure. }
\end{figure}

The $^{107}$Ag($\alpha,\gamma)^{111}$In and $^{107}$Ag($\alpha$,n)$^{110}$In reaction cross sections have been measured in the laboratory energies range between \mbox{8.16 MeV} and \mbox{13.00 MeV}, which includes a part of the astrophysically relevant energy range. Laboratory energies have been converted into effective center-of-mass energies ($E_\mathrm{c.m.}^\mathrm{eff}$) that correspond to beam energies in the target at which half of the yield of the full target thickness is obtained \cite{Rolfs87}. The experimental cross section results for $^{107}$Ag($\alpha,\gamma)^{111}$In and $^{107}$Ag($\alpha$,n)$^{110}$In reactions are presented in Tables \ref{tab:XS} and Fig. \ref{fig:ag} and \mbox{Fig. \ref{fig:an}}. Previous results from Baglin \cite{Baglin2005} and Stelson \cite{Stelson64} are also included in the figures. For $^{107}$Ag($\alpha,\gamma)^{111}$In reaction, disagreement with Baglin \cite{Baglin2005} is not understood, but the comparison with theory makes the Baglin values very unlikely. For $^{107}$Ag($\alpha$,n)$^{110}$In reaction, the agreement is good with Stelson \cite{Stelson64} but our energy range is much wider.

The uncertainty of the measured cross sections comprise the following partial components added quadratically: counting statistics (between \mbox{0.6$\%$} and \mbox{14.0$\%$}), detection efficiency (7$\%$) (including the conversion factor between the two counting geometries), decay parameters (less than \mbox{3.1$\%$)} and target thickness (\mbox{7$\%$}). The uncertainty of the beam energy is governed by the energy loss in the targets determined with the SRIM code \cite{SRIM} (between \mbox{0.6$\%$} and \mbox{1$\%$}), uncertainties in the energy degrader foil thickness (\mbox{1$\%$}) and the energy calibration and stability of the cyclotron (\mbox{0.5$\%$}). In order to check systematic uncertainties, measurements at \mbox{11 MeV} and \mbox{11.5 MeV} energies were carried out with two different targets. The cross section results of the two measurements are in a good agreement (Table \ref{tab:XS}).

The ($\alpha$,n) reactions on $^{107}$Ag populate the ground state (\mbox{$T_{1/2}$ = 4.92 h}) and isomeric state (\mbox{$T_{1/2}$ = 69.1 min}) of
$^{110}$In. Partial cross sections leading to these two states are listed separately in the Table \ref{tab:XS_an}. The total cross section of the $^{107}$Ag($\alpha$,n)$^{110}$In reaction was determined by summing the partial cross sections. In those cases where the cross section was determined based on the counting of more than one $\gamma$-line (see Table \ref{tab:decaypar}), the final cross section quoted in the tables and shown in the figures were obtained by weighted average. 

\begin{table}
\caption{Partial cross sections of the $^{107}$Ag($\alpha$,n) reaction leading to the ground and isomeric states of $^{110}$In.}
\begin{ruledtabular}
\begin{tabular}{lccc}
\multicolumn{1}{c}{E$_{\rm beam}$} & \multicolumn{1}{c}{E$_{\rm c.m.}^{eff}$} & \multicolumn{2}{c}{Cross Section [$\mu$b]} \\
\multicolumn{1}{c}{[MeV} & \multicolumn{1}{c} {[MeV]} & \multicolumn{1}{c}{$^{110g}$In (4.92 h) } & \multicolumn{1}{c} {$^{110m}$In (69.1 min)} \\
\hline

9.00\footnotemark[1]  & 8.57  $\pm$ 0.06 & 0.25 $\pm$ 0.03 & 7.3   $\pm$ 1.3  \\
9.50\footnotemark[1]  & 9.07  $\pm$ 0.06 & 2.8  $\pm$ 0.2  & 48.8  $\pm$ 4.4  \\
10.00\footnotemark[2] & 9.55  $\pm$ 0.09 & 15.7 $\pm$ 1.0  & 219   $\pm$ 16   \\
10.50\footnotemark[2] & 10.05 $\pm$ 0.09 & 58.2 $\pm$ 3.5  & 745   $\pm$ 55   \\
11.00\footnotemark[1] & 10.52 $\pm$ 0.07 & 170  $\pm$ 11   & 1821  $\pm$ 130  \\
11.00                 & 10.57 $\pm$ 0.07 & 175  $\pm$ 11   & 1883  $\pm$ 148  \\
11.50\footnotemark[1] & 11.00 $\pm$ 0.07 & 518  $\pm$ 31   & 4949  $\pm$ 352  \\
11.50                 & 11.04 $\pm$ 0.07 & 506  $\pm$ 31   & 4911  $\pm$ 363  \\
12.00                 & 11.51 $\pm$ 0.07 & 1494 $\pm$ 90   & 12443 $\pm$ 886  \\
12.50                 & 12.00 $\pm$ 0.07 & 2904 $\pm$ 176  & 21663 $\pm$ 1542 \\
13.00                 & 12.50 $\pm$ 0.08 & 4668 $\pm$ 281  & 32398 $\pm$ 2321 \\

\end{tabular} \label{tab:XS_an}
\end{ruledtabular}
\footnotetext[1]{measured with enriched target.}
\footnotetext[2]{measured with an energy degrader foil.}
\end{table}

\subsection{Comparison with Hauser-Feshbach predictions}
\label{sec:theory}

The Hauser-Feshbach model of compound nuclear reactions makes use of averaged widths describing particle or photon emission from the formed compound nucleus \cite{Raus00,HauF52,Raus11}. These averaged widths comprise sums over transition widths connecting the compound state and individual final states, determined by computing transmission coefficients from the solution of a time-independent Schr\"{o}dinger equation for each transition energetically possible and allowed by quantum mechanical selection rules \cite{HauF52,Raus11}. In addition to binding energies of the involved nuclei, optical potentials and low-lying, discrete excited states have to be known for the calculation of averaged particle-widths, and the $\gamma$-strength function, discrete excited states, and nuclear level density enter the computation of the $\gamma$-width.

\begin{figure}
\includegraphics[angle=-90,width=\columnwidth]{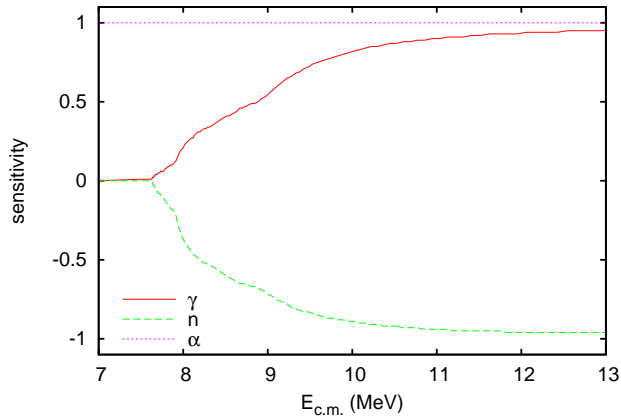}
\caption{\label{fig:sensi_ag} 
(Color online) Sensitivity of the $^{107}$Ag($\alpha$,$\gamma$)$^{111}$In reaction cross sections to variations in various averaged reaction widths as function of energy \cite{sensi}. The cross sections are insensitive to a variation of the proton width across the shown energy range.}
\end{figure}

For a correct interpretation of the differences between data and predictions, it is necessary to study the sensitivities of the cross sections to the calculated averaged widths which, in turn, depend on different nuclear properties. These sensitivities are not only different for different reaction types but they are also energy dependent and, in consequence, variations of certain nuclear properties may have different impact on the resulting cross sections at lower and higher energy. Sensitivities as a tool to understand the origin of discrepancies between data and theory have been thoroughly discussed in \cite{sensi} and have been used in previous investigations similar to the present one (e.g., see \cite{Halasz12,Kiss12,Raus12,Sauerwein12,Kiss14,Glo14,Net14,Kiss15,Gur15}).

\begin{figure}
\includegraphics[angle=-90,width=\columnwidth]{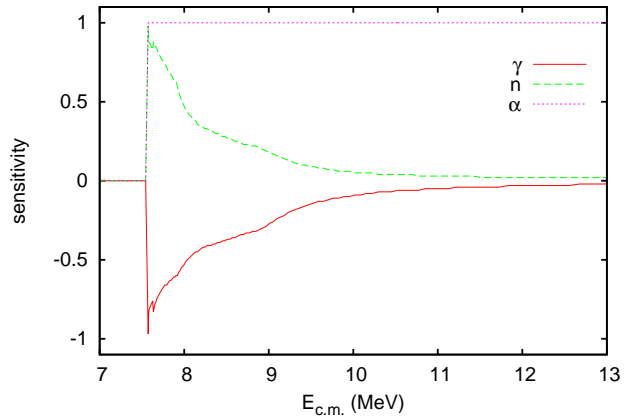}
\caption{\label{fig:sensi_an} 
(Color online) Same as Fig.\ \ref{fig:sensi_ag} but for $^{107}$Ag($\alpha$,n)$^{110}$In.}
\end{figure}

In general, the cross sections may be sensitive to several properties at a given energy. In this case, it is an advantage to have consistent data for two or more reaction channels at the same energy. Here, we are able to simultaneously consider ($\alpha$,$\gamma$) and ($\alpha$,n) data which allows to reduce ambiguities. The sensitivity factors of the cross sections of both reactions to variations in the averaged widths are shown in Figs.\ \ref{fig:sensi_ag} and \ref{fig:sensi_an}. A sensitivity factor $-1\leq s \leq 1$ implies the cross section changing by a factor
$f=\left|s\right|\left(v-1\right)+1$, when the corresponding width is changed by a factor of $v$ \cite{sensi,Raus14}. For $s\geq 0$, the original cross section has to be multiplied by $f$ whereas for $s<0$ it has to be divided by $f$. This means that a negative sensitivity shows that the cross section will change in the opposite direction than the width, i.e., it will increase when the width decreases and vice versa. As can be seen in Figs.\ \ref{fig:sensi_ag} and \ref{fig:sensi_an}, the cross sections of both reactions are sensitive to the $\alpha$ width in the same manner across the investigated energy range but the sensitivity to neutron- and $\gamma$-widths are different and opposite. Both reactions are insensitive to a change of the proton width at the shown energies.

It should be noted that astrophysically relevant energies are located below the ($\alpha$,n) threshold and therefore the astrophysically interesting width is the $\alpha$ width. This led to the series of investigations to better constrain this width at low energy, as mentioned in Sec.\ \ref{sec:intro}. It was found that the previous data could be described using an energy-dependent modification of the $\alpha$ width which only acts at low energy \cite{Som98,Sauerwein11,Raus12,Kiss14,Glo14,Kiss15}. The $\alpha$ width was calculated using the well-known optical potential by \cite{McF} with one modification: the depth of the volume imaginary part $W$ was made energy-dependent. It has to approach the value given in \cite{McF} (\mbox{25 MeV}) at high energy but has to be shallower at energies below the Coulomb barrier energy $E_\mathrm{C}$. A Fermi-type function was used to achieve this:

\begin{equation}
W=\frac{25}{1+e^{\left(0.9E_\mathrm{C}-E_\mathrm{c.m.}^\alpha \right)/a_E}} \quad \mathrm{MeV}.
\label{eq:param}
\end{equation}

In previous work, the value $a_E$ for the ``diffuseness'' of the Fermi-type function has been found to be between 2 and \mbox{5 MeV}, depending on the reaction. Using such a modified, effective optical potential it remains an open question whether the modification is really due to a required change in the optical potential, which affects the total reaction cross section, or due to the neglection of direct processes in the entrance channel \cite{Raus13}.

Here, we use a similar approach to be able to reproduce the experimental data. Figure \ref{fig:ag} and \ref{fig:an} compares calculations performed with the SMARAGD code \cite{SMARAGD} with the data. It can be seen that the prediction using the optical potential by \cite{McF} (labeled ``Theo'') follows the ($\alpha$,n) data quite well except at the lowest measured energy. On the other hand, the energy dependence of the ($\alpha$,$\gamma$) data is reproduced well but the calculation gives cross sections which are about $2-3$ times too large.

As found in Figs. \ref{fig:sensi_ag} and \ref{fig:sensi_an}, at the upper end of the measured range the ($\alpha$,n) reaction is only
sensitive to the $\alpha$ width. Since the data are reproduced at these energies, the $\alpha$ widths have to be accurately predicted there.
At the same energies the ($\alpha$,$\gamma$) reaction is sensitive not only to the $\alpha$ width but also to the $\gamma$- and neutron widths.
Since these widths have exactly opposite impact on the cross sections, only the change in the ratio $q=\Gamma_\gamma/\Gamma_\mathrm{n}$ of
average $\gamma$ width $\Gamma_\gamma$ to average neutron width $\Gamma_\mathrm{n}$ can be determined from the requirement to reproduce the ($\alpha$,$\gamma$) data simultaneously with the ($\alpha$,n) data. Rescaling $q$ by a factor of 0.5 shifts the predicted cross sections down and excellent agreement with the experimental ($\alpha$,$\gamma$) cross sections is achieved at the higher energies.

Even with the adjusted ratio $q$, cross sections at the lowest measured energies remain overpredicted. According to the sensitivities, the only way to mend this is to alter the $\alpha$ width. The $\alpha$ width, however, describes well the data at higher energies and therefore an energy-dependent modification is required. We chose the same parameterization as used in previous work and given in Eq. (\ref{eq:param}). We found that the best fit to the data can be obtained with \mbox{$a_E=5$ MeV}. The resulting excitation functions are also shown in Figure \ref{fig:ag} and \ref{fig:an} and labeled ``Theo (mod)''. These results are fully consistent with previous investigations, where a similar $a_E$ was found and a similar rescaling of the $\gamma$ width relative to the neutron width was necessary.

\section{Summary and Conclusion}
\label{sec:summary}

The $^{107}$Ag($\alpha,\gamma)^{111}$In and $^{107}$Ag($\alpha$,n)$^{110}$In reaction cross sections have been measured in the effective center of mass energies between \mbox{7.79 MeV} and \mbox{12.50 MeV}, with the aim to extend the available database for improving predictions of the averaged $\alpha$ widths at low energy.  Experimental results were compared with Hauser-Feshbach statistical model calculations. It was found that an energy-dependent modification of the $\alpha$ width and a rescaling of the $\gamma$- to neutron-width ratio is necessary. This is completely consistent with previous works. This finding confirms the applicability of the previously suggested parameterization of the optical $\alpha$+nucleus potential also at mass numbers lower than studied so far.

\begin{acknowledgments}

This work was partially supported by the Scientific and Technological Research Council of Turkey (TUBITAK), Grants No. 109T585 (under the EUROGENESIS research program) and Grants No. 108T508 and by OTKA grants K101328 and K108459. CY acknowledges support through the Scientific and Technological Research Council of Turkey (TUBITAK), under the programme of BIDEB-2219. GGK acknowledges support from the Bolyai grant. TR is supported by the BRIDGCE grant from the UK Science and Technology Facilities Council (grant ST/M000958/1), by the Swiss NSF, and the European Research Council (grant GA 321263-FISH).

\end{acknowledgments}

\end{document}